\documentclass[runningheads]{svmult}

\usepackage{makeidx}   
\usepackage{graphicx}  
\usepackage{subeqnar}  
\usepackage{multicol}  
\usepackage{physprbb}  
\makeindex             

\def\theta{\vartheta}
\def\phi{\varphi}
\def\sax{{\it Beppo}SAX}

\def\ltsima{$\; \buildrel < \over \sim \;$}
\def\lsim{\lower.5ex\hbox{\ltsima}}
\def\gtsima{$\; \buildrel > \over \sim \;$}
\def\gsim{\lower.5ex\hbox{\gtsima}}

\begin{document}
\title*{Failed optical afterglows}
%
%
%
%
\titlerunning{Failed optical afterglows}
%
\author{Gabriele Ghisellini\inst{1}
\and Davide Lazzati\inst{1,2} \and Stefano Covino\inst{1}}
\authorrunning{G. Ghisellini et al.}
%
%
\institute{Osservatorio Astronomico di Brera, Via Bianchi 46
I-23807 Merate (Lc), Italy \and Present Address: University
of Cambridge, Institute of Astronomy, Cambridge CB3 0HA, UK}

\maketitle              

\begin{abstract}
While all but one Gamma--Ray Bursts observed in the X--ray band 
showed an X--ray afterglow, about 60 per cent of them have not been detected 
in the optical band.
We show that this is not due to adverse observing conditions.
We then investigate the hypothesis that the failure of detecting 
the optical afterglow is due to absorption at the source location.
We find that this is a marginally viable interpretation, but only if the 
X--ray burst and afterglow emission and the possible optical/UV flash 
do not destroy the dust responsible for absorption in the optical band.
If dust is efficiently destroyed, we are led to conclude that bursts 
with no detected optical afterglow are intrinsically different.
\end{abstract}

\section{Observations}

Figure 1 shows magnitudes of the detected bursts 
and upper limits of failed optical afterglows (FOAs),
all in the $R$ band, versus the time of observation. 
Filled and empty circles correspond to \sax~and non--\sax~bursts with 
detected optical afterglows, while arrows are upper limits. 

\begin{figure}
\begin{center}
\includegraphics[width=1\textwidth]{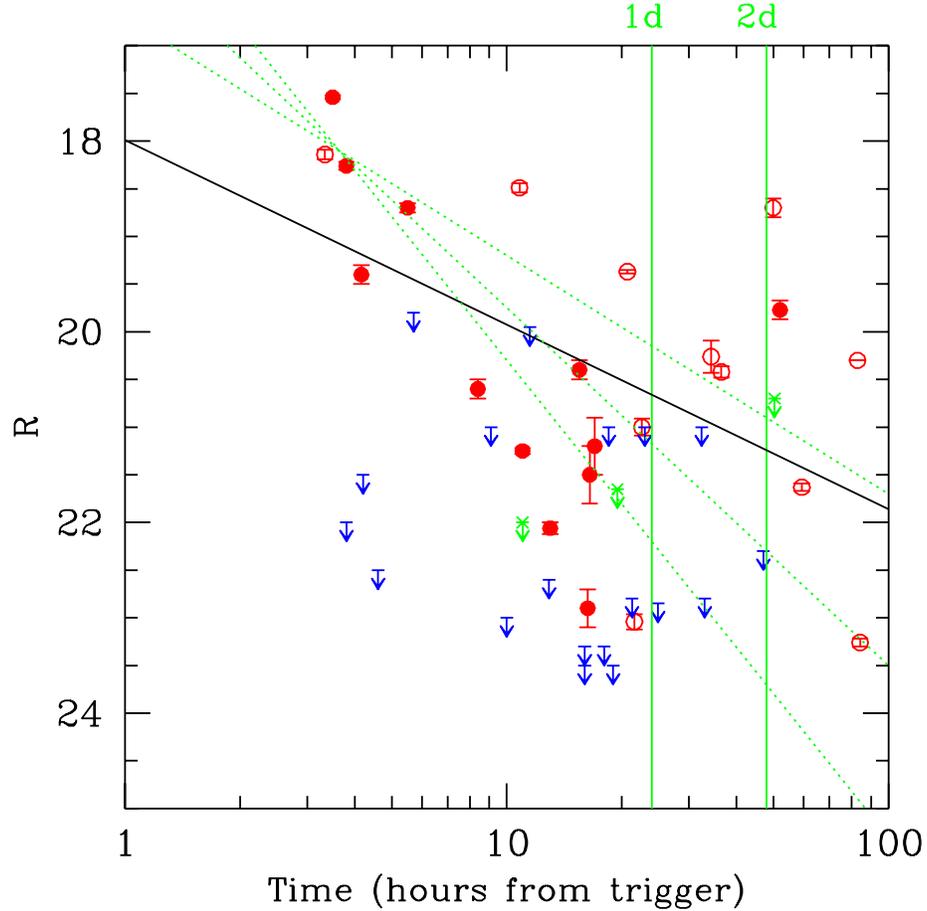}
\end{center}
\vskip -0.5 true cm
\caption[]{Detection $R$ magnitude (or upper limits) versus the time 
of observation for a set of afterglows. 
Filled circles show
optical detections of \sax~afterglows while empty circles
show detections of non \sax~afterglows.
Arrows show upper limits for \sax~failed optical afterglows.
Arrows with crosses refer to the upper limits on $\gamma$--ray poor
X--ray transients detected by \sax (their inclusion/exclusion from 
the sample does not alter any of the presented result).
The dark solid line is the best fit for the magnitudes of detections 
vs. time. Dotted lines show the $F_\nu(t)\propto t^{-1}$, 
$t^{-1.5}$ and $t^{-2}$ relations. From Lazzati et al., 2001.}
\label{fig:opt}
\end{figure}

The visual inspection of Figure~\ref{fig:opt} reveals a clear segregation
of arrows from dots, the former being systematically fainter than the
latter at comparable times. This impression is confirmed by the 
application of a bidimensional KS test (Press et al. 1992). 
The probability for the circles (empty $+$ filled)
and the arrows being derived from the same parent distribution 
is $P\sim$ 0.2 per cent.

This result shows that in most cases we failed to detect the optical 
afterglow not because the search was conducted without the necessary depth, 
but instead because the FOAs are indeed fainter than the detected ones. 
Yet, it is possible that FOAs are optically fainter because 
intrinsically less energetics at all wavelengths, or because they
are more distant. 
In order to check this, we compared the X--ray and $R$ band flux 
densities of bursts with and without optical detection 12~hours after 
the burst event, finding that
the X--ray fluxes of FOAs are not systematically 
fainter than the fluxes of afterglows with optical detection, 
indicating that FOAs are indeed optically poor and define a 
different population with respect to optically detected afterglows.

We have checked that local Galactic extinction does not play a crucial role 
by comparing the hydrogen column densities in the direction of detected 
afterglows with those in the direction of FOAs. 

\section{Intrinsic Absorption?}

We have investigated the possibility that the difference between 
the two groups is due to absorption local to the burst.
We can quantify in roughly 2 magnitudes the amount of 
average absorption in the $R$ band needed for more than half of the bursts
to go undetected in the optical.

Can a typical molecular cloud produce such an absorption in more than half
of the bursts?
In order to answer this question we (Lazzati et al. 2001) have computed
the {\it average} (i.e. over many line of sights) 
and the {\it maximum} absorption of known molecular
clouds in our galaxy, taking into account that observations
of bursts in the $R$ band 
actually correspond to light emitted at shorter wavelengths, where
extinction is more effective.
Since the redshifts of FOAs is obviously unknown, we have assumed 
$z=1$ for all of them.

Our results can be summarized as follows:

\begin{itemize}
\item If the burst is located at random within a molecular cloud,
it will on average be absorbed with the mean value of the cloud absorption.
In this case we found that only a few percent of the burst afterglows
could be missed for this reason.

\item If bursts are located in star forming regions, then they lie
in the densest parts of the cloud, i.e. those with maximum absorption.
In this case it is (albeit marginally)
possible that up to 60 per cent of the bursts
have optical afterglows sufficiently absorbed to have avoided detection.
But consider that we have been very conservative in our procedure, because 
our results are based on considering {\it upper limits} on the optical
flux, and {\it peak} absorption columns expected in giant molecular clouds.

\item The latter assumptions may well be too conservative, if the dust is
bound to evaporate when illuminated and heated by the powerful 
optical/UV flash of the gamma--ray burst (Waxman \& Draine 2000) and
by its X--ray radiation (Fruchter et al. 2000).
This dust sublimation is suggested for a sample of burst afterglows 
(Vreeswijk et al. 1999, Galama \& Wijers 2000), in which a very large 
hydrogen column density $N_{\rm H} \gsim 10^{22}$ cm$^{-2}$,
as estimated by X--ray data, is associated with almost no optical extinction. 
The results can be understood only in terms of a dust to gas ratio 
$\sim 100$ times smaller than the Galactic average value. 
In turns, such low values of the dust to gas ratio can be explained only 
if the dust has been completely sublimated in the surroundings of the burst.
Indeed the theoretical models mentioned above predict that dust can be
destroyed by the burst emission out to a radius comparable to the dimension 
of a typical molecular cloud (up to a few tens of parsecs).
If this is the case, the material responsible for absorption in FOAs 
is not the overdense cocoon surrounding the star forming region, 
but the cloud as a whole (or even less), and the discrepancy between 
the observed and measured value 
becomes extremely compelling.

\end{itemize}

\end{document}